\documentclass[final,3p,times]{elsarticle}

\usepackage{amsmath}

\usepackage{amssymb}

\journal{Statistical Mechanics and its Applications Physica A}

\begin{document}

\begin{frontmatter}



\title{Scaling in Income Inequalities and its Dynamical Origin}


\author[label1]{Zolt\'an N\'eda} 
\author[label1]{Istv\'an Gere}
\author[label2]{Tam\'as S.~Bir\'o}
\author[label3]{G\'eza Toth}
\author[label4]{Noemi Derzsy}

\address[label1]{Babe\c{s}-Bolyai University, Dept. of Physics, Cluj-Napoca, Romania}
\address[label2]{Wigner RCP, Budapest, Hungary and Complex Science Hub, MedUni Vienna, Austria}
\address[label3]{Central Statistical Office of Hungary, Budapest, Hungary}
\address[label4]{AT\&T Labs, Data Science \& AI Research,  New York,  USA}

\begin{abstract}
We provide an analytically treatable model that describes in a unified manner income distribution for all income categories.
The approach is based on a master equation with growth and reset terms. The model assumptions on the growth and reset rates are tested on an exhaustive database with incomes on individual level spanning a nine year period in the Cluj county (Romania).  In agreement with our theoretical predictions we find that 
income distributions computed for several years collapse on a master-curve when a properly normalised income is considered. The Beta Prime distribution is appropriate to fit the collapsed data and it is shown that distributions derived for other countries are following similar trends with different fit parameters. 
The non-universal feature of the fit parameters suggests that for a more realistic modelling the model parameters have to be linked with specific socio-economic 
regulations.
\end{abstract}




\end{frontmatter}







\section{Introduction}
Starting with the seminal work of Vilfredo Pareto \cite{Pareto} the wealth and income distribution in a 
given society have been in the focus of social science studies \cite{piketty}. 
Borrowing tools from statistical physics, econo-physics \cite{econophysics} is also keenly interested in this problematics. 
By using modern data-mining  and data-processing techniques together with various modelling approaches econo-physicists observed 
and confirmed many universalities in wealth and income distribution \cite{chakraborti}. 
Ranging from simple analytically solvable mean-field type models, to more elaborated agent-based or network-based 
computational models many techniques are used for explaining the statistics of the observed data \cite{yakovenko}.

Investigating experimentally the income distribution is simpler than investigating the wealth distribution. 
For income there are already many accurate, exhaustive and electronically available data (social security, tax, etc.)
\cite{ishikawa,dragulescu,clementi,silva,figueira,souma_2001,richmond,derzsy,soares_2016,abreu_2019}.
Wealth is usually estimated in a non-direct manner, using some quantities that are believed to be 
a proxy for it \cite{dragulescu,sinha,abul-magd,hegyi}. 
Although wealth and income are different quantities, their distribution presents several similarities \cite{chakraborti,yakovenko}. 
The famous $80-20$ Pareto law is valid for both of them: on average $20\%$ of the population owns $80\%$ of the total wealth 
or alternatively $20\%$ of the population gets $80\%$ of the total income in a given society. 
This law is a consequence of the largely stretched wealth and income distribution, which is proved to have a "heavy", 
power-law type tail. The exponent characterizing the power-law decay of the cumulative distribution function is known 
as the Pareto exponent. Apart from this well documented stylised fact, it is also known that in the low and middle income 
classes the distribution functions of both wealth and income are nearly Boltzmannian, 
i.e. they follow an exponential trend \cite{atkinson_2000,dragulescu_2001}. 

In the realm of models the situation is opposite. 
Most of the econo-physics models are targeting the wealth distribution 
\cite{bouchard_2000,solomon_2001,pianegonda_2003,iglesias_2003,scafetta_2004,chatterjee_2004,garlaschelli_2004,sinha_2005,coelho_2005,grupta_2006,gade_2007,ausloos_2007,gade_2009,berman_2015} 
and there are rather few modelling attempts for the income. 
The models developed for income usually consider a combination of additive and multiplicative processes as stochastic effects 
for the dynamics of salaries \cite{silva,milakovic_2005,yakovenko_2009_springer}. 
Both for wealth and income distributions {\em the exponential and power-law regions are modeled separately}. 
Due to the fact that the Boltzmannian exponential distribution is a standard equilibrium distribution, 
the key challenge for physicists is to understand the scaling regime. 
Up to our knowledge momentarily there is no model that i) is based on proven  socio-economic assumptions
and ii) is successful in describing the whole income or wealth interval in a unified manner. 
The present study intends to fill this niche.

Our aim here is quite ambitious: starting from exhaustive, long-term and individual level income data \cite{derzsy} 
we derive average trends that are used as input for a general master-equation \cite{Biro-Neda}.
From here we derive the stationary distribution that can fit the whole income interval. 
We demonstrate that income distributions computed from a complete social dataset from Romania and Hungary 
are in excellent agreement with our model results and rescale on a general master curve.
That curve fits to a very simple functional form.  
The rest of the paper is then organised as follows: (section 2)  the growth and reset master-equation is briefly reviewed; 
(section 3) the analysed data are described and experimental trends for the averaged dynamics of the individuals' income are derived; 
(section 4) the observed trends are used in the framework of the growth and reset model and from these trends 
the expected stationary income distribution is computed; 
(section 5) scaling properties and universal features of the income distribution are revealed; 
(section 6) the income dynamics is discussed in the view of the observed growth and reset rates and experimental distributions for income are  compared
critically with model prediction; 
(section 7) final conclusions are drawn.


\section{The growth and reset process}
\newcommand{\be}{ \begin{equation} }
\newcommand{\ee}[1]{\label{#1} \end{equation} }

\begin{figure}[!ht]
    \centering
		\includegraphics[width=9cm]{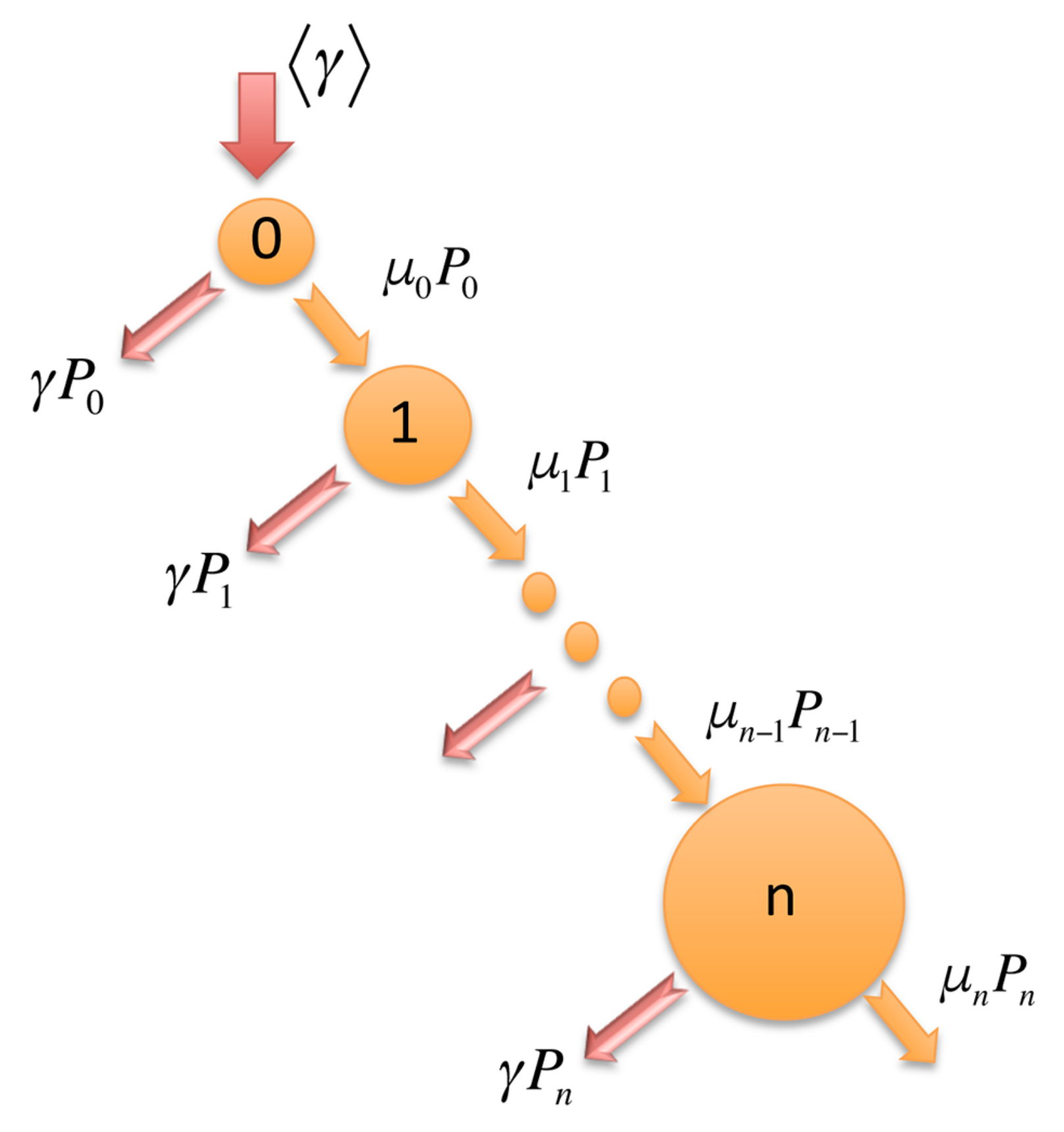}	
	\caption{Schematic illustration for the 'growth and reset' process in the probability space.}
	\label{growth-reset}
\end{figure}

Recently a simple master equation was considered,
containing both local and long distance transitions: uni-directional one-step growth and reset to zero terms \cite{Biro-Neda}. 
For a brief review, let us consider a system composed of many identical elements that can have different 
numbers of quanta  of a relevant quantity. 
For example, this can be a human society composed of a large number of individuals with different amounts of wealth or income.  
We denote by $P_n(t)$ the probability that an element has exactly $n$ quanta at time $t$. Normalisation requires $\sum_n P_n(t)=1$.  
The flow diagram for the growth and reset process in the probability space is presented in Figure \ref{growth-reset}, 
while the  dynamical evolution equation reads as

\begin{equation}
\frac{dP_n(t)}{dt}=\mu_{n-1}P_{n-1}(t)-\mu_n P_n(t)-\gamma_nP_n(t) + \delta_{n,0}\langle \gamma \rangle(t) . 
\label{master_dis}
\end{equation} 
We denote the growth-rate by $\mu_n$ and the reset rate $\gamma_n$ from the state with $n$ quanta.

\newcommand{\ba}{\begin{eqnarray}}
\newcommand{\ea}[1]{\label{#1} \end{eqnarray} }
\newcommand{\nl}{\nonumber \\}

The first and second term on the right hand side are due to one-step-growth, 
the third term stands for the contribution from the reset to zero. The last term is 
a feeding term at state zero, necessary in order to preserve normalisation. 
The $\langle \gamma \rangle$ value can be calculated from the normalisation condition:
\begin{equation}
\langle \gamma \rangle(t) =\sum_{j} \gamma_j P_j(t).
\end{equation} 
One easily generalises the above description for the case when continuously labelled states are considered. 
Instead of the discrete state index $n$ we introduce continuous variables, $x$,
and the discrete $P_n(t)$ probabilities will be replaced by a continuous $\rho(x,t)$ probability density 
function, satisfying the normalisation condition $\int_{\{x\}} \rho(x,t) dx=1$. 
Inspecting now a discretisation with bin length $\Delta x$ we define  our model master equation 
in terms of continuous variables as:
\begin{equation}
 \begin{split}
 P_n(t)\rightarrow \rho(n\Delta x,t) \Delta x &= \rho(x,t) \Delta x \\
 \mu(x)=\mu(n\Delta x) &= \mu_n \Delta x \\
 \gamma(x)=\gamma(n\Delta x) &= \gamma_n.
\end{split}
\end{equation}

The continuous limit of the master equation (\ref{master_dis}) becomes

 \begin{equation}
 \frac{\partial \rho(x,t)}{\partial t}=-\frac{\partial}{\partial x} \left[ \mu(x) \rho(x,t) \right] - \gamma(x) \rho(x,t) +\langle \gamma(x) \rangle (t) \delta(x),
 \label{master_gen}
 \end{equation}
 with:
 \begin{equation}
 \langle \gamma(x) \rangle (t) = \int_{0}^{\infty} \gamma(x) \rho(x,t) dx
 \label{normal}
 \end{equation} 
 In the above equation we denoted by $\delta(x)$ the Dirac functional.  It has been proven, that under very general conditions the above dynamical evolution equation converges to a steady-state with a $\rho_s(x)$ stationary probability density \cite{BTN}. 
This stationary probability density is derived from the condition:
 \begin{equation}
 -\frac{\partial}{\partial x} \left[ \mu(x) \rho_s(x) \right] - \gamma(x) \rho_s(x) +\langle \gamma(x) \rangle \delta(x) \: = \: 0,
 \label{master_gen2}
 \end{equation} 
 with:
 \begin{equation}
 \langle \gamma(x) \rangle = \int_{0}^{\infty} \gamma(x) \rho_s(x) dx
 \label{normalx}
 \end{equation} 
 
\newcommand{\eadx}[1]{{\rm e}^{#1} }
Its solution is given in the following analytical form:
 \begin{equation}
\rho_s(x) \: = \: \frac{\mu_0 \rho_s(0)}{\mu(x)} \, \eadx{-\int_0^x\limits \frac{\gamma(u)}{\mu(u)}du}.
\label{stat-distr}
 \end{equation}
As we have pointed out in \cite{Biro-Neda} a plethora of important distributions, 
that are frequently encountered in complex systems, can be generated by
properly selecting the local growth rate $\mu(x)$ and the reset to zero rate $\gamma(x)$.


\section{Experimental data}

In order to study the dynamics of income in a properly delimited social system 
we use a long-term exhaustive dataset from Cluj County (Romania) \cite{derzsy}.
The anonymized dataset provides monthly income information for each employee (whose identity has been encrypted) between 2002 and 2009.  Assuming that the growth and reset model is applicable here, our aim is to gather experimental information 
on the shape of the $\gamma(x)$ and $\mu(x)$ kernel functions. 

The dynamics in the growth of the salaries can be estimated by following the change in the yearly average salary 
of each employee in two consecutive years. If the average salary of employee 
$i$ in year $k$ is $w_i(k)$, we determine the $\Delta w_i(k)=w_i(k+1)-w_i(k)$ quantity. 
We group the employees in exponentially increasing bins (logarithmic binning) based on their average salary in year $k$. 
This means that in bin $j$ we will have those employees, whose salary is between $2^j$ and $2^{j+1}$. In each bin
we then determine the average change in the salary: 
\begin{equation}
\langle \Delta_j w(k)\rangle=\langle \Delta w_i(k) \rangle_{\{i|w_i(k)\in[2^j,2^{j+1}]\}}. 
\end{equation}
In order to improve the statistics, we perform an averaging on the years too. We obtain the
\begin{equation}
\langle \Delta_j w \rangle=\langle \Delta_j w(k) \rangle_k
\end{equation}
time averaged  values. In the average we have excluded the first complete year of the database 
(2002) and the years after 2007, when the economic crisis hit Romania and rearranged 
the salaries in a drastic manner. The data for the years 2008 and 2009 are irrelevant for a long-term trend since 
in year 2008 for avoiding state-bankruptcy all salaries in the public sector were 
cut by 25\%, massive dismissal of public servants was implemented and other drastic financial and social-economic measures were taken. 

\begin{figure}[!ht]
	\centering
		\includegraphics[width=12cm]{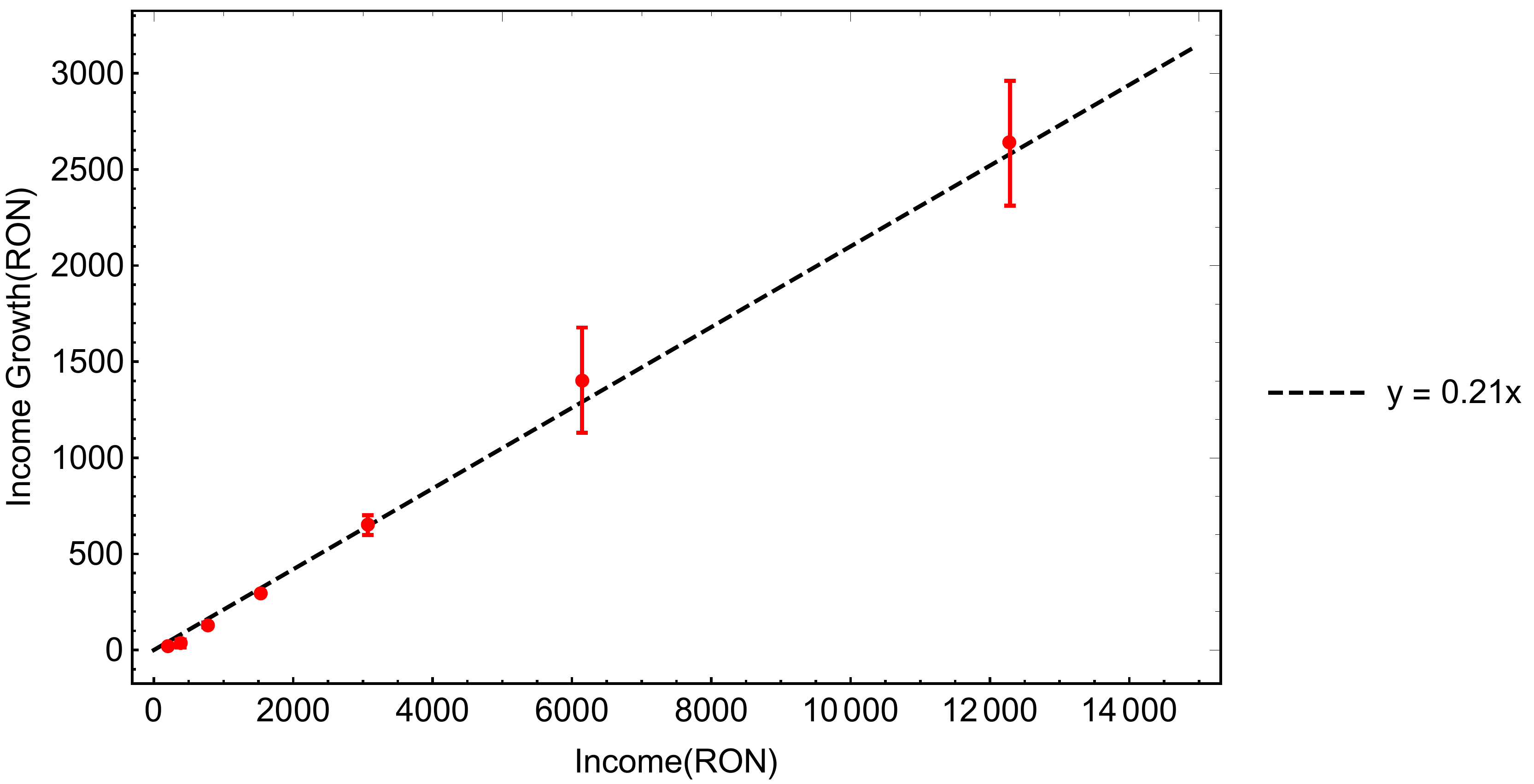}	
\caption{Averaged growth in salary  ($\langle \Delta_j w\rangle$) as a function of the salary $w_j$ ($w_j=3/2\cdot2^j$). 
The dashed line is a linear fit with the proportionality constant: $C=0.21$. The goodness
of the fit is characterized by the $R^2=0.986$ correlation value.}
	\label{growth-rate}
\end{figure}

The $\langle \Delta_j w \rangle$  values plotted for each bin against $w_j=2^j*3/2$ 
(the middle value of the bin $j$ for the income) leads  to the trend presented in Figure \ref{growth-rate}. For the plotted data 
we indicate by error bars the standard deviation of the data for different years. 
We have plotted the data only up to approximately ten times the average salary, 
since above this value the statistics become poor due to the small number of employees in the given bin. 
The data strongly suggest a linear trend, i.e.  $\langle \Delta_j w\rangle =C \cdot w_j$, 
in agreement with previous findings in Japan \cite{aoyama_2003,fujiwara_2003}. 
As it will be discussed in the next section, this multiplicative growth  implies a linear kernel function 
for the growth-rate: $\mu(x)=\beta x$. 
 
\begin{figure}[!ht] 
	\centering
		\includegraphics[width=10cm]{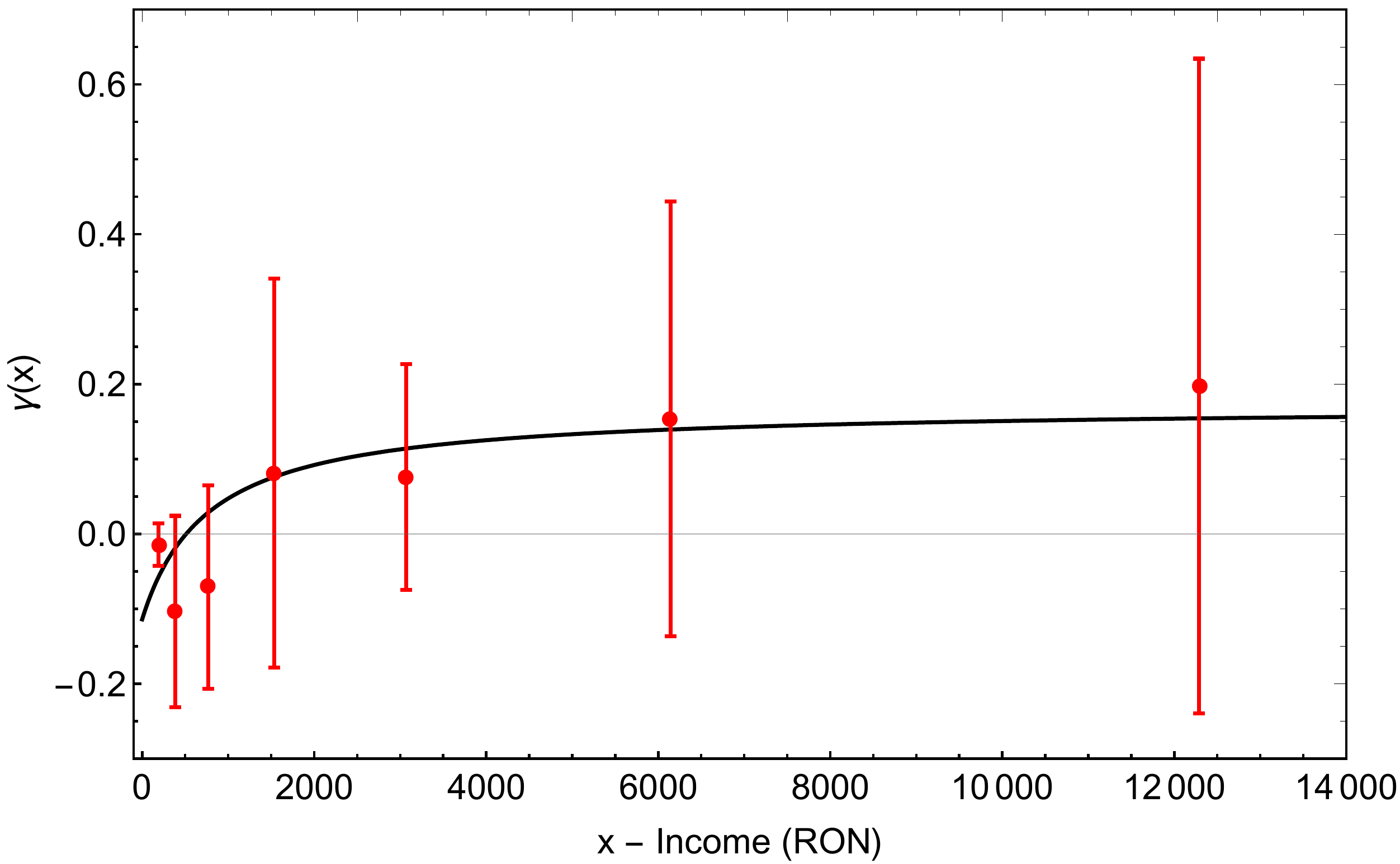}	
\caption{The averaged reset rate for bin $j$, $\langle \gamma_j \rangle$, as a function of the mean salary in the bin, $w_j$. 
The continuous line represents  the fit: $3\beta [1-5\langle x \rangle/(3x+3 \langle x \rangle)]$, ( justified in section 4.), 
with $\beta=0.057$, an optimally chosen fit parameter, and the average salary calculated for the 2003-2007 period 
$\langle x \rangle=767$ RON.}
	\label{reset-rate}
\end{figure}

For gathering information on the reset rate, $\gamma(x)$, one has to identify in each year the persons that are newly 
appearing in the system and those who are leaving it. For a given income interval  $[w,w+\Delta w]$ the reset rate is 
by definition positive, if more workers are leaving the system from this income interval than entering it. 
On the other hand, the reset rate is negative if more workers appear in the database than leaves it. 
By definition, a worker is leaving the system if it was continuosly present until year $k$, and it is not present in year $k+1$ 
and thereafter. Similarly, a worker is appearing in the database if it was not present until year $k$, and it is present in 
year $k$ and after.  Let us denote by $N^{out}_j(k)$ the number of those workers that are leaving the system in year $k$ 
with a salary that is in bin $j$ (i.e. their last salary $w(k)$ has the property: $w(k)\in[2^j,2^{j+1}]$), 
and by $N^{in}_j(k)$ the number of workers that are entering in year $k$ with a salary  that is in bin $j$. 
We define the reset rate for bin $j$ as: 
\begin{equation}
\gamma_j(k)=\frac{N^{out}_j(k)-N^{in}_j(k)}{N_j(k-1)} 
\end{equation} 
where $N_j(k-1)$ is the number of workers in year $k-1$ with salary in bin $j$ ($w(k-1)\in [2^j,2^{j+1}]$). 
Again, similar to the growth rate, in order to achieve a better statistics,  one can 
average over all years $k$ and obtain: $\langle \gamma_j \rangle=\langle \gamma_j(k) \rangle_k$.  
For the averaging we have used the years from 2003 up to 2007, for the same reason as the one stated for the growth-rate. 
Again, the data were plotted only up to approximately ten times the average salary, where the statistics is reasonably good. 
The values of $\langle \gamma_j \rangle$ as a function of $w_j$ are plotted in Figure  \ref{reset-rate} and we indicate by error bars 
the standard deviation of the data for different years. The data points outline the shape of the $\gamma(x)$ reset rate kernel: it is negative for small income values and positive 
for larger income. This trend is the one that one would naturally expect: new workers (young people) enter the system 
with predominantly smaller salaries while old people retire with larger salaries. 
In such a view the low income region is dominated by the incoming workers (the average reset rate is negative), 
while the larger income categories have a net outflow of workers (reset rate is positive).

We will argue in the next sections that in view of the observed income distribution, the reset rate should have the form:
\begin{equation}
\gamma(x)=3\beta \left( 1-\frac{5\langle x \rangle}{3(x+\langle x \rangle)} \right).
\label{gammakernel}
\end{equation}  
 
Here $\langle x \rangle$ denotes the average salary, which was $767$ RON for the period between 2003 and 2007 
in Cluj county. The value of $\beta$ is a fitting parameter. 
On Figure \ref{reset-rate} we display a certain fit to the reset kernel function  (\ref{gammakernel}), choosing $\beta=0.057$.
Due to the large error bars many other fitting forms are suitable to describe the observed trend. To motivate the chosen form
we recall that for a linear growth rate only a few reset rate kernels will lead to analytical solution in the probability density (\ref{stat-distr}). 
Even if one finds such other kernel that is also compatible with the trend, the problem remains to obtain consistent parameters 
with a good fit for the probability density function of incomes.

\section{Income distribution from the perspective of
 the growth and reset model}
 
The dynamics behind income (salary) differs from person to person. 
For modelling purposes we consider an average trend for the salaries,
and approach the income distribution in the framework of our growth and reset model using average growth and
reset rates. In this sense our model is a mean-field approach. 

The model, described in previous sections, is realistic for the dynamics in the salaries in the sense that it
assumes that the salaries of individuals predominantly do show an increasing trend. 
The reset process appears when a person retires or a new worker enters in the considered social system. 
Senior people retire and young ones appear in the system. A positive reset rate means leaving the system, 
while a negative reset rate belongs to a new appearing in the studied ensemble. 
New workers appear mostly with low salaries, in the lower income categories. 
Retiring is mostly from the higher income categories, therefore one should use {\em smart reset rates} that reproduce this basic 
feature. 
The data derived from the ten year social security survey in  Cluj county (Romania) (section 3) 
confirm the above expectations and is compatible with the following form of the kernel function (see Figure \ref{reset-rate}):
\begin{equation}
\gamma(x)=K-\frac{b}{x+q}
\label{resetrate}
\end{equation} 
Here $K$, $b$, and $q$ are all freely adjustable parameters. For small income values this leads to a negative reset rate, 
while in the limit of large income it becomes positive, saturating at the value $K$. 

\begin{figure}[!ht]
	\centering
		\includegraphics[width=8cm]{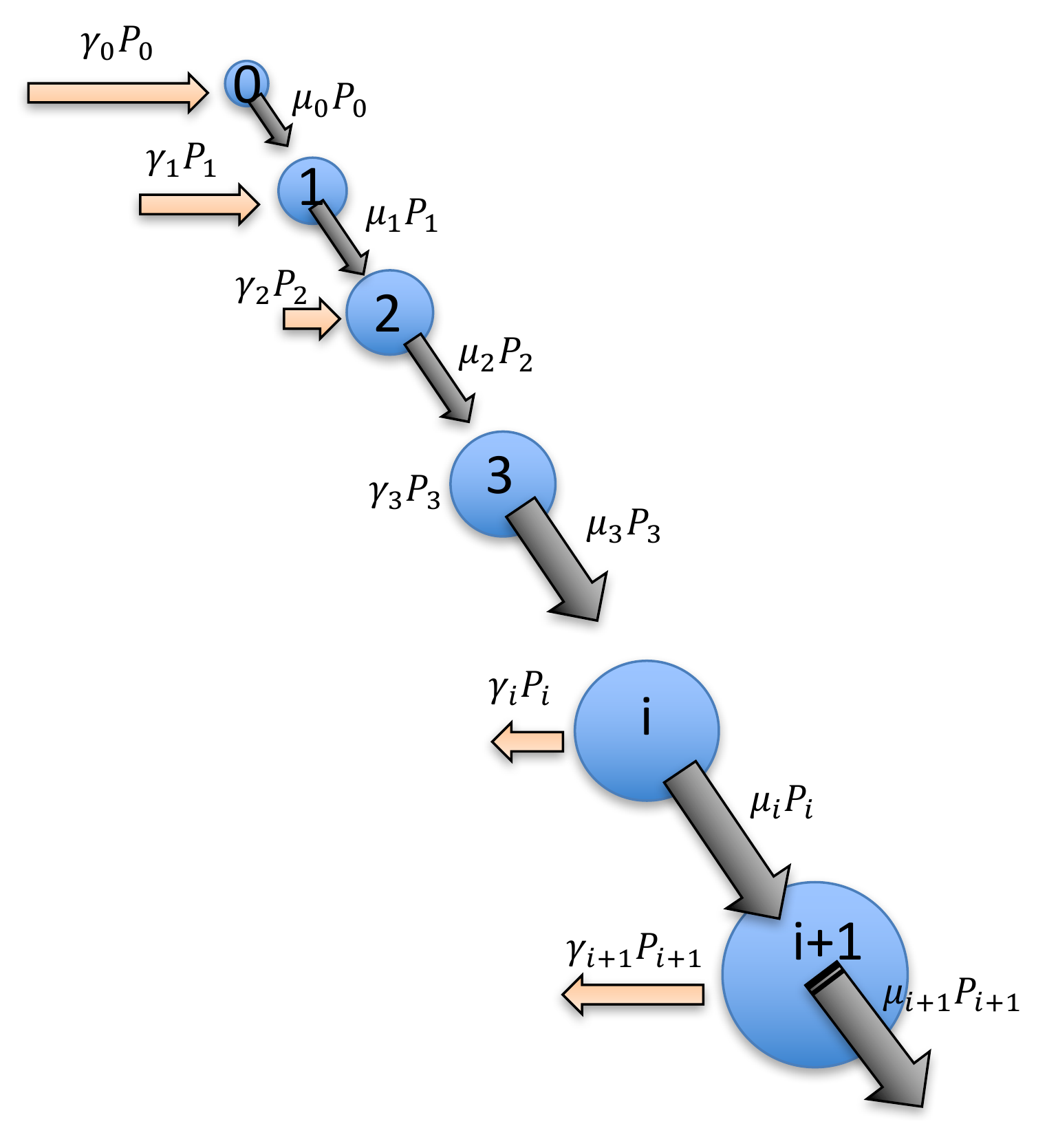}	
\caption{Schematic illustration of the growth and reset process for the income distribution with realistic growth and reset rates.}
	\label{smart-reset}
\end{figure}

Salaries tend to increase with a given percentage rather than with a fixed amount. 
Inflation also leads to such a multiplicative increase.  The data derived from the real-world social system presented 
in the previous section (Figure \ref{growth-rate})
suggest a {\em linear preferential growth}. According to this, in a time interval $T$ the growth of salary $x_i$ is given as: 
\begin{equation}
\Delta_T( x_i) = x_i(t+T)-x_i(t)=u \cdot T \cdot (x_i+g),
\end{equation}
where  $u$ and $g$ are two constants. 
Assuming that both the growth rate and the income are continuous variables, the growth speed is
best modeled by:
\begin{equation} 
\frac{dx}{dt}=\frac{\Delta_T(x)}{T}= u \cdot (x+g).
\end{equation}
The average time, $\tau(x)$, needed for a $dx$ growth in salary is inversely proportional to the growth speed. The growth rate 
$\mu(x)$ used in the growth and reset model (\ref{master_gen}) is on the other hand inversely proportional to $\tau(x)$. 
According to these assumptions we use
\begin{equation}
\mu(x)=\beta\cdot (x+g),
\label{pref-growth}
\end{equation}
where $\beta$ is yet another proportionality constant. 
The flow with our smart-reset rate (\ref{resetrate}) and preferential growth-rate (\ref{pref-growth}) 
is illustrated schematically for the discrete probability space in Figure \ref{smart-reset}. 

The stationary distribution $\rho_s(w)$ (\ref{stat-distr}) for the growth and reset rates
(\ref{pref-growth}) and (\ref{resetrate}) is obtained as a Pearson type I distribution,

\begin{equation}
 \rho_s(x) \: \sim \: \left( x+ q \right)^{-\frac{b}{\beta(q-g)}} \,
 \left(x +g \right)^{\frac{b}{\beta(q-g)}-\frac{K}{\beta}-1}.
\end{equation}

In the limit $g\rightarrow 0$ (realistic for the income dynamics plotted in Figure \ref{growth-rate}) 
one reconstructs the familiar Beta prime distribution:

\begin{equation}
 \rho_s(x)=  q^{\left(\frac{K}{\beta}-\frac{b}{\beta q}\right)} \frac{\Gamma \left( \frac{b}{q \beta} \right)}{\Gamma \left( \frac{b}{q \beta}-\frac{K}{\beta} \right) \Gamma \left( \frac{K}{\beta} \right)}  
 \left( 1+ \frac{x}{q} \right)^{-\frac{b}{\beta q}} \,
 x ^{\frac{b}{\beta q}-\frac{K}{\beta}-1}.
\label{betaprime}
\end{equation}

The first moment of this distribution, the medium income appears as
\begin{equation}
\langle x \rangle=\int_0^{\infty} x \rho_s(x) dx= q\frac{\left( \frac{b}{\beta q}-\frac{K}{\beta} \right) }{\left( \frac{K}{\beta}-1 \right)}.
\end{equation}

Rescaling now the income relative to the mean, we arrive at the form

\begin{equation}
\langle x \rangle \rho_s(x) = \left( \frac{a-s }{s-1} \right) ^{a-s} \frac{\Gamma(a)}{\Gamma(a-s) \Gamma(s)}\left( 1+ \frac{x}{\langle x \rangle} \frac{a-s}{s-1} \right)^{-a} \,
\left( \frac{x}{\langle x \rangle} \right) ^{a-s-1},
\label{beta-prime-scaled}
\end{equation}

where we have used the notations $a=b/\beta q$ and $s=K/\beta$.


\section{Scaling in the experimental income distributions}

\begin{figure}[!ht]
	\centering
		\includegraphics[width=15cm]{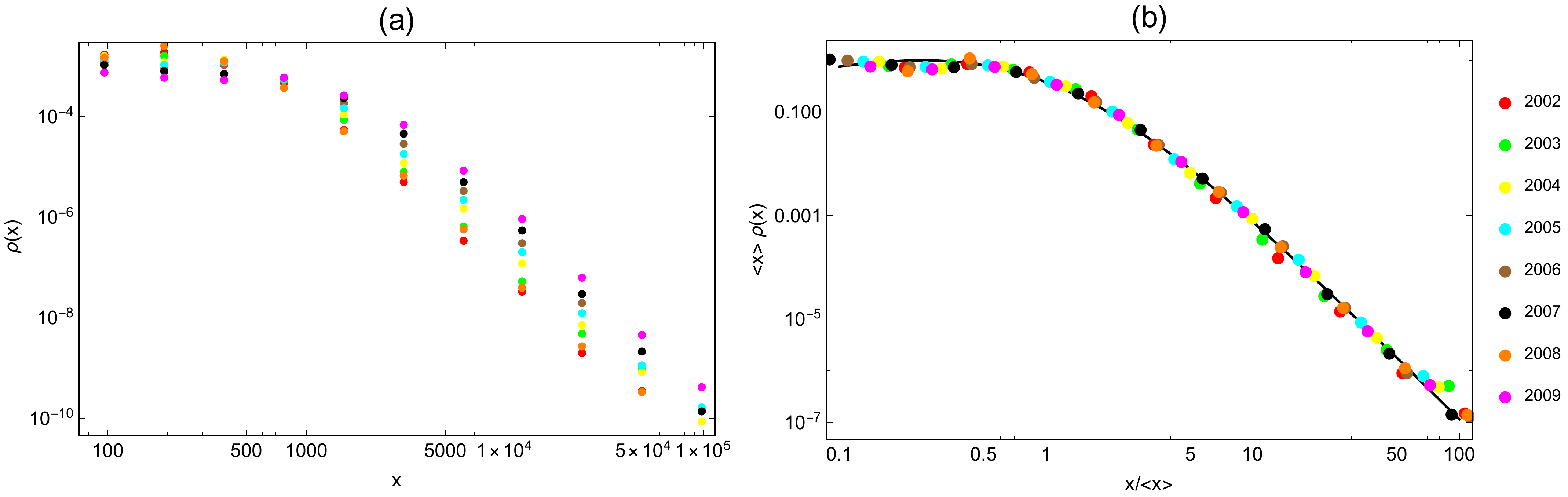}	
	\caption{ (a) Normalised income distribution density functions for the Cluj county (Romania) in different years.
	(b) Collapsed data when the axis are properly rescaled.
	The continuous line indicates a fit of the form $f(x)=12x(1+x)^{-5}$. Monthly salaries are considered and given in RON. 
	Data point with different colours are results for different years as indicated in the legend. }
	\label{distr-rough}
\end{figure}

The normalised income distribution function $\rho(x)$ was computed for several databases. 
First, the exhaustive data obtained from the social security 
records for  Cluj county (Romania) \cite{derzsy} was considered for each year from 2002 up to 2009. 
The distribution functions were obtained by a logarithmic binning method, grouping the salaries in income bins so that the bin sizes were increased
exponentially. The obtained distribution functions are plotted on a log-log scale in Figure \ref{distr-rough}a and one can readily observe the expected Pareto-like tail. 

Moreover, one will observe that these curves collapse, when plotting $\langle x \rangle \rho_{(x)}$ as a function of $x/ \langle x \rangle$ 
(Figure \ref{distr-rough}b). The fit of the form (\ref{beta-prime-scaled}) describes excellently the collapsed data, 
if one chooses $a-s-1=1$ and $a=5$. In such a case one uses $s=3$ leading to the master-curve:
\begin{equation}
\langle x \rangle \rho_s(x)= 12 \frac{x}{\langle x \rangle} \left( 1+\frac{x}{\langle x \rangle} \right)^{-5}
\label{master-curve}
\end{equation}

\begin{figure}[!ht]
	\centering
		\includegraphics[width=11cm]{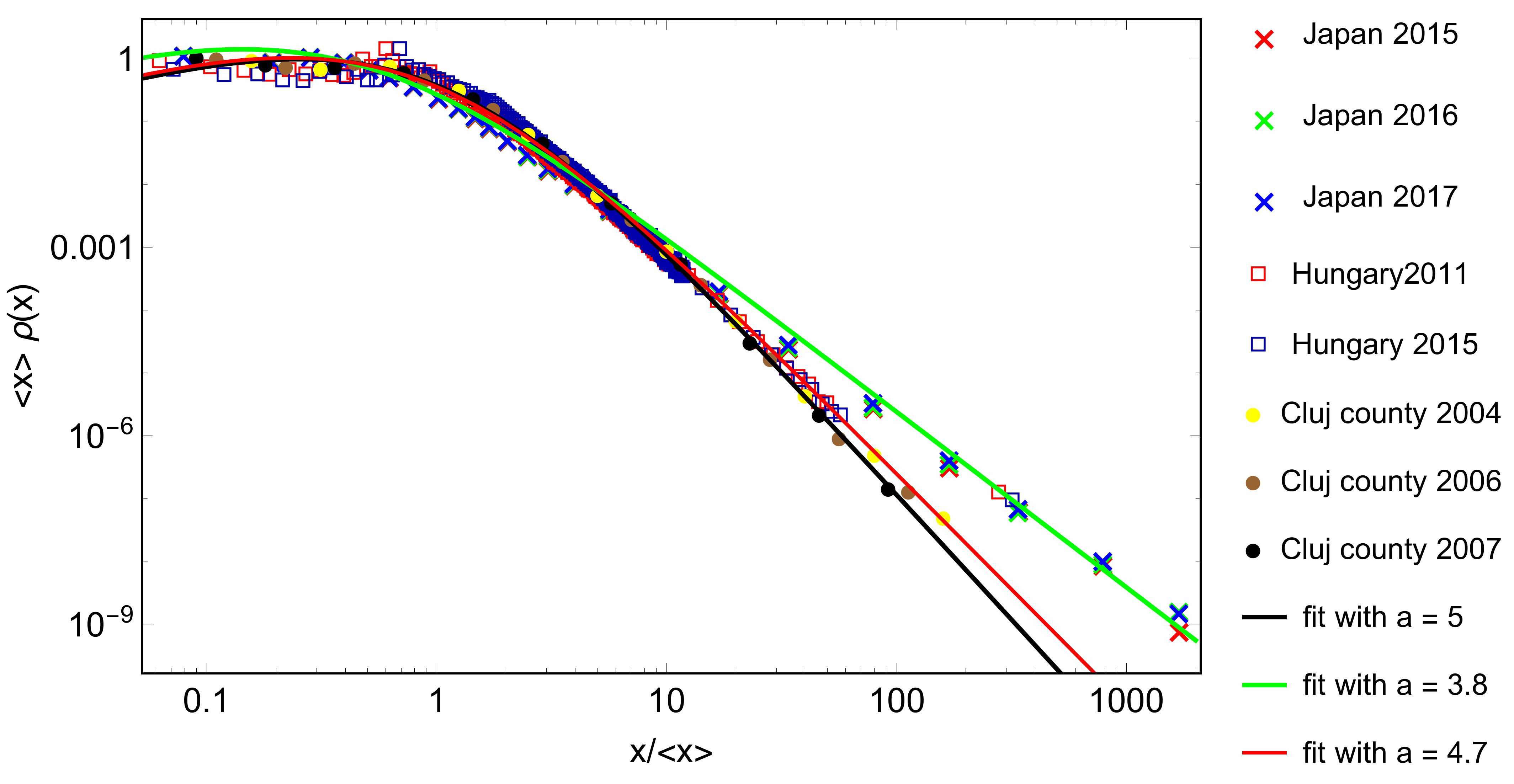}	
	\caption{Rescaled income distribution for different countries and regions.
	The continuous lines indicates fit of the form  (\ref{beta-prime-scaled}) with $s=a-2$ and $a=5$ for Cluj county, $a=4.7$ for Hungary and $a=3.8$ in the case of Japan.}
	\label{highres-data}
\end{figure}

In order to verify whether there is a much deeper universality in the rescaled data bridging over different countries, taxation and social security systems, 
we have considered income distribution for several other countries as well. High resolution data derived from taxation was obtained for Hungary (from the Central 
Statistical Office of Hungary)  for the year 2011 and 2015, and for Japan \cite{Japan-data} over several years (2011-2017). On such high resolution data the power-law tail of the distribution is clearly visible. On Figure \ref{highres-data} we plot together the data for Cluj county, Hungary and Japan and fit all of them with the 
Beta prime distribution (\ref{beta-prime-scaled}), similarly with (\ref{master-curve}), considering the only fitting parameter $a$ and fixing $s=a-2$. By this the simple power-law term in the Beta prime distribution will have the exponent $1$. The data from Figure \ref{highres-data} suggests that the $a$ parameter is 
clearly not uniform and depends on the chosen country. For Hungary the best fit parameter is $a=4.7$, while for Japan we find $a=3.8$ in comparison with the value for the Cluj county  $a=5$.  In case of the same country however, similarly with what we obtained in Romania, the data for different years rescale.

\begin{figure}[!ht]
	\centering
		\includegraphics[width=11cm]{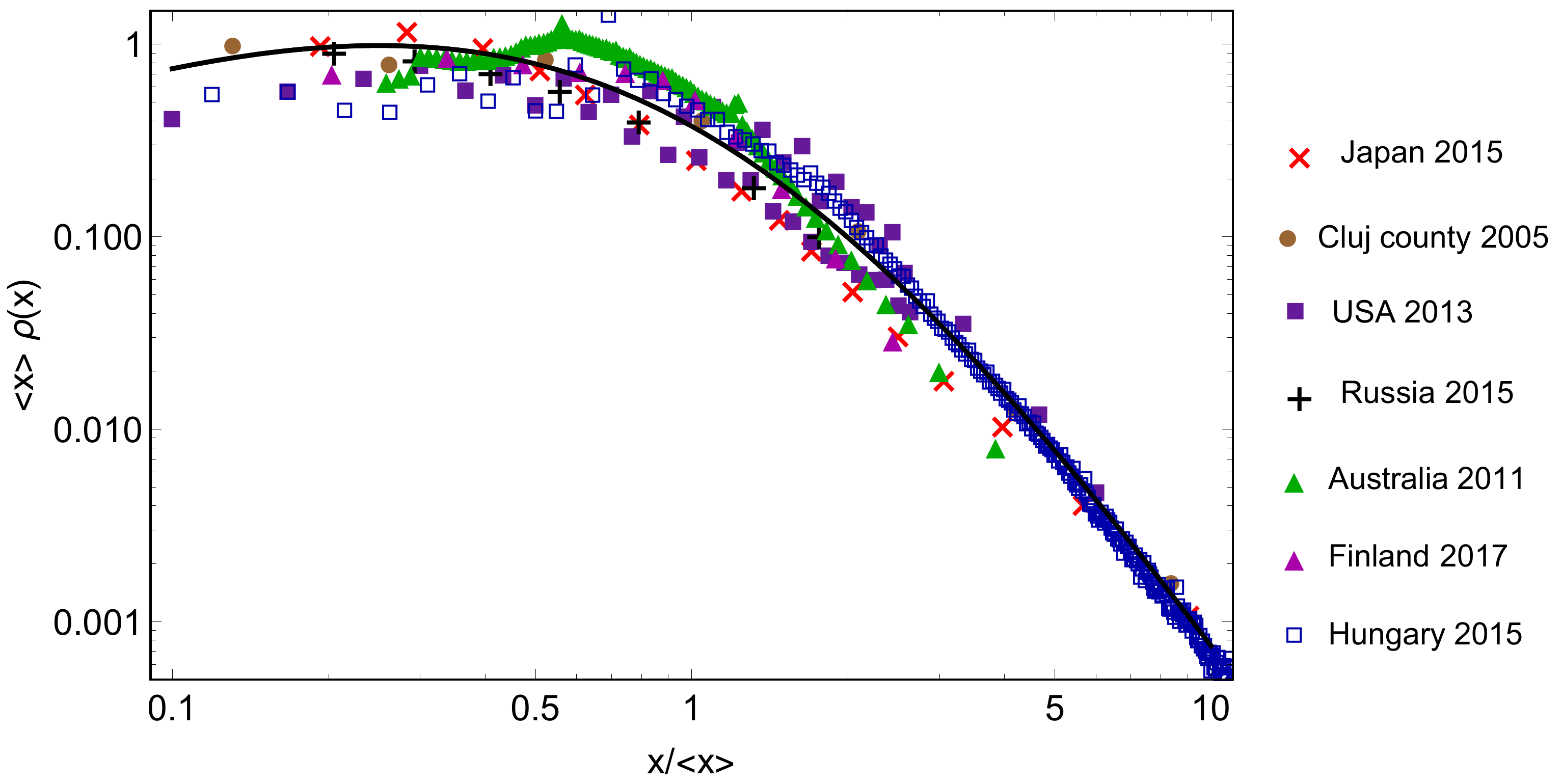}	
	\caption{Rescaled income distribution for different countries and regions.
	The continuous line indicates a fit of the form $f(x)=12x(1+x)^{-5}$}
	\label{scaled-distr}
\end{figure}

Much lower resolution data, up to $x/\langle x \rangle=10$, is available for many other countries. Considering data from Australia (2011)  \cite{Australia-data}, Finland (2017)  \cite{Finland-data},  and census survey data from USA (2013)  \cite{USA-data} and Russia (2016) \cite{Russia-data} we show a comparison with the data from Cluj county, Japan and Hungary on Figure \ref{scaled-distr}. Due to the fact that on such lower resolution data we do not see the region
of the rich, one gets the false impression that  the trends are rather similar. A more closer look on Figure \ref{scaled-distr} reveals however that the collapse is far from being acceptable.

\section{Discussion}

In order to verify now the dynamical hypothesis inherent in the growth and reset model 
we return to the assumed reset and growth rates, 
equations (\ref{resetrate}) and (\ref{pref-growth}).  
As it was already presented in Figure \ref{growth-rate} the linear preferential 
growth in the mean salaries is nicely observable in the experimental data. 
The dynamic data suggest also $g=0$, so the simplification  used in arriving at the Beta Prime 
distribution (\ref{betaprime}) is justified. 

Taking $a=5$, $s=3$ and imposing that the total individuals in the system remains constant without the input given by the $\delta (x)$ Dirac functional term at $x=0$

\begin{equation}
\Delta N_{total} =\langle \gamma(x) \rangle=\int_0^{\infty} \gamma(x) \rho_s(x) dx=\int_0^{\infty} \left( K-\frac{b}{x+q}\right)  12 \frac{x}{\langle x \rangle^2} \left( 1+\frac{x}{\langle x \rangle} \right)^{-5} dx=0,
\end{equation}
one will get:

\begin{equation}
 \begin{split}
q &=\langle x \rangle  \\
K &=3 \beta \\
b &=5 \beta \langle x \rangle
\end{split}
\end{equation}

For the reset rate kernel (\ref{resetrate}) the above constraints lead us to the form used in (\ref{gammakernel}), 
providing a reasonable fit for the experimentally determined trend (Figure \ref{reset-rate}).

It is easy to show that the growth and reset model preserves the total amount of income in the steady state.   
In our particular case it is immediate to show that the net change is:
\begin{equation}
\Delta W_{total} =\int_0^{\infty} (\mu(x)-x \gamma(x))\rho_s(x)=\int_0^{\infty} \left( \beta x  - 3 \beta x \left (1 -\frac{5\langle x \rangle}{3(x+\langle x \rangle)} \right)  \right)12 \frac{x}{\langle x \rangle^2} \left( 1+\frac{x}{\langle x \rangle} \right)^{-5}dx=0
\end{equation}

The fact that both the total number of workers and the total amount of income are stationary in the framework of our model 
implies that the average salary should also be roughly constant. 

We perform now an overview for the Cluj data, and discuss this in the framework of our model. 
In Table \ref{table-cluj}, we present the values of the total number of registered 
workers and the average income for each year.  
Here the variation of the total number of workers as a function of time shows no clear trend. 
The average salary, however, is steadily  increasing except the year 2008, when the Romanian economy 
had been strongly affected by the world economic crisis. 

\begin{table}[h]
\caption{Overview of the total number of workers and average salaries for the Cluj county database.}
\label{table-cluj}
\begin{center}
\begin{tabular}{ |p{1.5cm}||p{1cm}|p{1cm}|p{1cm}|p{1cm}|p{1cm}|p{1cm}|p{1cm}|p{1cm}|p{1cm}| }
 \hline
 \bf{year}  & 2002 & 2003 & 2004 & 2005 & 2006 & 2007 & 2008 & 2009 \\
 \hline
 \hline
 $N_{total}$  & 179136 & 189095 & 182964 & 188363 & 197021 & 217267 & 186580 & 264209 \\
 \hline
 $\langle x \rangle (RON)$ &  461 & 550 & 615 & 732 & 873 & 1068 & 447 & 1359 \\
 \hline
\end{tabular}
\end{center}
\end{table}

The data presented in Table \ref{table-cluj} reveals however, that the average salary apparently increases in time, 
except in year 2008 when the economic crisis showed its effect in the Romanian economy. 
The reason for this apparent paradox is that the model does not take into account the total economic growth and inflation, 
which just properly re-scales all incomes and prices. This rescaling effect is the one that would explain the increase 
in the average salaries, and it is not captured by the equilibrium solution on $x/\langle x \rangle$ 
in the growth and reset model. 

We return and comment now on the incomplete collapse of the data plotted in Figure \ref{scaled-distr} and the differences in the $a$ fitting parameter 
for different countries. These results suggests 
that country specific economic differences are important in understanding and modelling income inequalities. Simple physical models based on
oversimplified growth and reset mechanism with universal rates (the approach considered here) might be successful at understanding rough trends and 
coarse-grained shapes of the income distribution function. For a complete description however, one has to go beyond these simple models and introduce country-specific 
socio-economic rules to understand the fine details of the distribution function. The present approach is successful in giving a unified big picture for the income distribution 
function and reveals a universal shape for the entire distribution function. 

Finally, let us comment on the power-law like tail of the distribution function. For most of the  countries the used data is not detailed 
enough to allow studying the scaling properties. Scaled incomes ($x/\langle x \rangle$) larger than 5 or 10 are all grouped together in one bin, with no specified upper bound. 
For Japan,  Hungary and Cluj county however we have have a fine binning in the high income limit as well, and this is the reason we do observe the Pareto-like tail. Interestingly, the rescaled data
for Hungary and Cluj county collapse in  reasonable manner (Figure \ref{highres-data}).  One can thus speculate that a reason for this is the quite similar 
socio-economic background, both countries being part of the former East-European block.  The much lower value of $a$ for Hungary and Romania, is an agreement with their socialist history. For Cluj county and Hungary we also have some indications that in the very high income limit ($x/ \langle x \rangle >100$) a second Pareto tail develops with a smaller Pareto exponent. 
This is in agreement with the already known fact  that the "very rich" are different, and the tail has yet another scaling \cite{veryrich}.  This second Pareto regime is however, not captured by our simple model. This is yet another reason why
for a more complete understanding of the income inequalities more complex models are still necessary.


\section{Conclusion}

We conclude that a simple master equation with state dependent growth and reset terms  \cite{Biro-Neda,BTN} 
performs beautifully in describing the income distribution on the whole range of income values. 
In agreement with the simple reasoning, the analyses of real-world data revealed that the growth 
in salaries is on average preferential, and the reset rate depends in a simple form as a function of income, 
being negative for the low income region and saturates at a constant positive value for high salaries. Experimental data for several years in the Cluj county (Romania) are in good agreement with the model assumptions and prediction. The rescaled distribution functions for different years collapse 
on a single master-curve, which has the right trend for income distributions derived in other countries as well.  
Our model predicts a Beta Prime distribution for the income and a simple scaling when the income is 
normalised relative to the mean value. Data for other countries confirm this findings. From the present study we learn that the much debated scaling-exponent for the 
Pareto-like tail should be determined using this fitting function 
instead of imposing an ad-hoc cut. The  trends observed in the income distribution function and in the 
mean growth and reset rates should motivate further studies and should be tested in other 
social systems in the future.

\section*{Author contribution}
Conceptualisation, research design and model was realised by Z.N and T.B. The experimental data was gathered by N.D, I.G and G.T.
Data processing was done by I.G and Z.N. First draft of the manuscript is by Z.N and T.B and all authors contributed to the final
version of the manuscript.

\section*{Acknowledgments}
Work supported by the UEFISCDI research grant PN-III-P4-ID-PCCF-2016-0084 and by the Hungarian National Innovation 
and Technology Office grant NKFIH-123815.
  
\section*{Disclosure}   

The present work does not reflect the position and opinion of AT\&T Labs and the Central Statistical Office of Hungary, it is solely the involved authors' view on the studied phenomenon.

\end{document}